\pdfoutput=1
\RequirePackage{ifpdf}
\ifpdf 
\documentclass[pdftex]{sigma}
\else
\documentclass{sigma}
\fi

\usepackage{DLdef1}

\def\smat#1{\left(\begin{smallmatrix}#1\end{smallmatrix}\right)}

\begin{document}

\allowdisplaybreaks

\newcommand{\arXivNumber}{2102.04192}

\renewcommand{\PaperNumber}{085}

\FirstPageHeading

\ShortArticleName{On a~Hidden Supersymmetry of Cosmological Billiards}

\ArticleName{On a~Hidden Supersymmetry\\ of Cosmological Billiards}

\Author{Dimitry LEITES~$^{\rm a}$ and Oleksandr LOZHECHNYK~$^{\rm b}$}

\AuthorNameForHeading{D.~Leites and O.~Lozhechnyk}

\Address{$^{\rm a)}$~Department of Mathematics, Stockholm University,\\
\phantom{$^{\rm a)}$}~Albanov\"agen 28, SE-114 19 Stockholm, Sweden}
\EmailD{\mail{dimleites@gmail.com}}

\Address{$^{\rm b)}$~Faculty of Mechanics and Mathematics, Taras Shevchenko National
University of Kyiv,\\
\phantom{$^{\rm b)}$}~Hlushkova Avenue 4e, 03127 Kyiv, Ukraine}
\EmailD{\mail{alozhechnik@gmail.com}}

\ArticleDates{Received September 16, 2024, in final form August 21, 2026; Published online August 29, 2026}

\Abstract{In the Belinski--Khalatnikov--Lifshitz cosmological models with an oscillatory approximation to singularity, the tracks of cosmological billiards (whatever they are) are realized in the Weyl chambers of the hyperbolic Lie algebras. The latter were classified by Li Wang Lai; their super analogs --- the almost affine Lie superalgebras --- were classified in arXiv:0906.1860. Here, we observe that some of the 142 hyperbolic Lie algebras $H$ can be ``superized'' to almost affine Lie superalgebras $S(H)$: for each of these $H$, it is possible to divide a row of its Cartan matrix by~2, simultaneously changing the parity of the corresponding Chevalley generators and relations between them. For the $H$ with a~symmetrizable Cartan matrix, we list all 97 such pairs ($H\longleftrightarrow S(H)$). Several (18 of the total 66) thus ``superizable'' algebras $H$ have multiple such ``superizations''. The tracks of cosmological billiards corresponding to both terms of these pairs ($H\longleftrightarrow S(H)$) coincide since $H$ and $S(H)$ have one and the same Weyl chamber. We also classify ``superizations'' of hyperbolic Lie algebras with non-symmetrizable Cartan matrix, and pairs $\mathfrak{g}\longleftrightarrow \mathfrak{g}_{\bar 0}$, where $\mathfrak{g}$ is a simple finite-dimensional Lie superalgebra without Cartan matrix and simple component~$\mathfrak{g}_{\bar 0}$.}


\Classification{17B50; 83E50; 17B20; 83F05}

\section{Introduction}

\subsection[Cosmological billiards and hyperbolic Lie algebras. Two hidden supersymmetries]{Cosmological billiards and hyperbolic Lie algebras.\\ Two hidden supersymmetries}

Studies of cosmological models
revealed a~wonderful, breath-taking connection between hyperbolic Lie algebras with symmetrizable Cartan matrices and the cosmological models demonstrating an oscillatory approximation to singularity, called Belinski--Khalatnikov--Lifshitz dynamics, see \cite{KhK1, BKhL}. Namely, the authors of \cite{KhK1, BKhL} had realized the tracks of cosmological billiards, whatever they are, inside the Weyl chambers of the hyperbolic Lie algebras.

Recently, a~hidden supersymmetry of these models was observed, see \cite{DH,DHJN,DHN, DSp, KKN}.

Here, we exhibit a completely different hidden ``supersymmetry'' of 66 cosmological billiards out of the total of 142 billiards corresponding to symmetrizable Cartan matrices of the hyperbolic Lie algebras. Namely, there are 97 ``superizations,'' 18 Cartan matrices (and hence, cosmological billiards) admit several ``superizations'' each, see~list \eqref{mult}. The tracks of the billiards corresponding to these pairs (Cartan matrix$\longleftrightarrow$its ``superization'') coincide since the Lie algebra and any of its ``superizations'' have one and the same Weyl chamber, see Section~\ref{opq}.
Since, apparently, the ``supersymmetry'' (``superizations'') we noted has nothing in common with the supersymmetry observed in \cite{DH,DHJN,DHN, DSp, KKN}, we do not overburden the present elementary note by recalling the astounding observation \cite{KhK1, BKhL}, see a review \cite{BH}, or the essence of \cite{DH,DHJN,DHN, DSp, KKN}.

\subsection{Hyperbolic Lie algebras and almost affine Lie superalgebras} In what follows, the ground field is $\Cee$;
all Cartan matrices $A$ are supposed to be indecomposable and normalized so that $A_{ii} = 0$ or 1 or 2; moreover, $A_{ij} =0\Longleftrightarrow A_{ji} =0$. For details and the construction of the Lie (super)algebra $\fg(A)$
from its Chevalley generators, see
\cite{BGL1, BLS, CCLL, GL}.

A Lie algebra $\fg(A)$ that grows polynomially in its Chevalley generators, is not finite-di\-men\-sional, and such that $A$ is indecomposable is called an \textit{affine Kac--Moody
algebra}.

To better visualize Lie \textit{super}algebras with Cartan matrix and of polynomial growth (unlike the non-super case, not all of them are affine Kac--Moody: some of them are ``stringy'', see \cite{CCLL}), we need the following relatively recently distinguished notion which elucidates which ingredients in the structure of affine Kac--Moody algebras are important and for what. Recall (for examples, see \cite{BKLS, BLS}) that a \textit{NIS-Lie $($super$)$algebra} $\fa$ is an algebra endowed with a Non-degenerate invariant symmetric bilinear form; and its \textit{double extension} is the Lie (super)algebra $\widehat \fa=(\fc\subplus\fa)\subplus\fd$, the semi-direct sum, where $\fc$ is a 1-dimensional center and $\fd$ is a 1-dimensional space of derivations. The double extension is ``interesting'', i.e., not isomorphic to the direct sum $\fa\oplus (\fc\oplus\fd)$, if the central extension is non-trivial and the derivation is outer, see \cite{BLS}.

The affine Kac--Moody Lie algebra is a~\textit{double extension} of $\fa^{\ell(1)}:=\fa\otimes\smash{\Cee\bigl[t, t^{-1}\bigr]}$, the loop algebra with values in a~simple finite-dimensional Lie algebra $\fa$, or of the ``twisted'' loop algebra~\smash{$\fa_\varphi^{\ell(k)}$} corresponding to the fixed points of the degree-$k$ automorphism $\varphi$ of the target Lie algebra~$\fa$; this~$\varphi$ corresponds to a~symmetry of the Dynkin graph of $\fa$. Each of these affine Kac--Moody algebras has a~symmetrizable Cartan matrix, and its Dynkin graph is the extended Dynkin diagram, see~\cite{K}.
\begin{center}
\begin{minipage}[l]{14cm}
In the super case, there appear new phenomena: there are simple finite-dimensional superalgebras $\fb$ without Cartan matrices, such that a~double extension of $\fb_\varphi^{\ell(k)}$ may have a~Cartan matrix; the other way round: even if $\fb$ has a~Cartan matrix, some of the twisted loop algebras $\fb_\varphi^{\ell(k)}$, or their double extensions, have no Cartan matrix; the NIS on $\fb$ can be either even or odd, see \cite{BLS}.
\end{minipage}
\end{center}

Therefore, the definition of affine Kac--Moody Lie algebra cannot be literally applied to Lie superalgebras $\fg$ of polynomial growth; we use the term
\textit{affine Kac--Moody Lie superalgebras} only for algebras $\fg$ with Cartan matrix, while algebras $\fg$ without Cartan matrix will be called
\textit{affine Lie superalgebras}.

For the results of classifications of Lie superalgebras with \textit{symmetrizable} Cartan matrix, see \cite{Se3,vdL}, where finite-dimensional and infinite-dimensional Lie superalgebras of polynomial growth are considered, and \cite{HS} for Lie superalgebras of polynomial growth with \textit{non-symmetrizable} Cartan matrices. A~general conjecture on the list of simple $\Zee$-graded Lie superalgebras of polynomial growth (see \cite{BLS, LSS, LSh}) analogous to Kac's conjecture proved by O.~Mathieu (see \cite{M}) is still open.

In the indescribable sea of Lie (super)algebras of exponential growth with Cartan matrix, there are islands consisting of interesting algebras; moreover, they can be completely classified. We will consider two such sets.

$\bullet$ A~given \textit{Cartan matrix} $A$ with entries in the ground field is said to be \textit{almost affine} if
the Lie (super)algebra determined by it is not finite-dimensional or affine Kac--Moody, but the Lie (super)algebra determined by any \textit{main} submatrix of $A$, i.e., the submatrix obtained by striking out any row and any column intersecting on the main diagonal, is a~sum of finite-dimensional or affine Kac--Moody Lie (super)algebras.

$\bullet$ A~given \textit{Lie $($super$)$algebra} $\fg(A)$
is called \textit{almost affine} if \textit{all its Cartan matrices are almost affine}.
Note that a~given Lie (super)algebra, though not almost affine, may have an almost affine Cartan matrix which goes into a~non-almost-affine matrix under an odd reflection. We are interested in the properties of a~given Lie (super)algebra, not in those of one or several of its Cartan matrices. Ignoring the difference between the two problems (classification of Cartan matrices vs. classification of Lie (super)algebras) have twice led to publications with mistakes, corrected in~\cite{CCLL}.

Clearly, there are many hyperbolic Lie algebras with $2\times 2$ Cartan matrices since their off-diagonal elements of the Cartan matrix can be arbitrary non-positive integers (according to the usual very restrictive convention, sometimes naturally broken, e.g., by Borcherds) or --- speaking about almost affine Lie superalgebras --- elements of the ground field, so such Lie (super)algebras are not considered in the literature, and neither do we. The Dynkin graph of an affine Kac--Moody (super)algebra is called an \textit{extension} of the Dynkin graph of a~finite-dimensional Lie (super)algebra, and the Dynkin graph of any hyperbolic Lie algebra is, obviously, an extension of an extended Dynkin graph. These extensions are obtained by very specific processes clearly defined only for Cartan matrices, not just by adding a node to the graph; the graph, in particular, the ``extended'' one, is just a~graphic illustration of the Cartan matrix useful (for humans) only if the matrix entries are integer, not very big in absolute value, etc. This is lucidly explained in \cite{CCLL}, together with bits of history of the subject, compare with \cite{DD}, where it is not even stated that the ``algebraically closed field'' of the scenery is of characteristic~$0$ (for characteristic $p>0$, see \cite{BGL, BLS}). The almost affine Lie algebra (not superalgebra) is usually called \textit{hyperbolic} (for a~reason, see \cite{CCLL}), or ``over extended'' meaning its Dynkin graph is an extension of an extended Dynkin graph.

Li Wang Lai was the first to classify all hyperbolic Lie algebras, see \cite{Li}. He gave his answer in terms of analogs of Dynkin graphs, but he left it to the reader to guess how his graphs correspond to Cartan matrices, and the method of getting the answer. Earlier, Kobayashi and Morita classified hyperbolic Lie algebras with symmetrizable Cartan matrix, see \cite{KM}. These results for symmetrizable matrices were double-checked in \cite{BS}, where the incomplete but very interesting paper \cite{S} was corrected; for the classification of all, in particular, non-symmetrizable, Cartan matrices of Lie algebras reproducing Li Wang Lai's result, see the arXiv version of \cite{CCLL}.
In \cite{CCLL}, almost affine Lie superalgebras are classified; the answer is given in terms of Cartan matrices, defined transparently and correctly together with analogs of Dynkin graphs.

\subsection{Non-uniqueness of the ``superization'' procedure}\label{sssSU} In this note, we intend to find out which of the hyperbolic Lie algebras allow a~``superization'' of the same type as $\fo(2n+1)$ does when it turns into $\fosp(1|2n)$. This ``superization'' consists of two steps: (i) divide, if possible without residue, a row by 2 as, for example, one divides the bottom row of the Cartan matrix of $\fo(2n+1)$ to get the Cartan matrix of $\fosp(1|2n)$
\begin{equation}\label{step1}
0\dots 0\ \ -2 \ 2 \longleftrightarrow 0\dots 0\ \ -1\ 1
\end{equation}
(ii) simultaneously, one has to change the parities (even $\longleftrightarrow $ odd) of the corresponding Chevalley generators and accordingly modify the relations between these generators, see~\cite{GL}.

Actually, it is this step (ii) which is crucial and leads to one of the two possible changes of the Cartan matrix: the first change is described in \eqref{step1}, the other one leads to the change $2 \longleftrightarrow 0$ of the diagonal element of the Cartan matrix together with the corresponding changes of the off-diagonal elements on the same row, see \cite{BLLS}, where these ``superizations'' are applied to classify (finite-dimensional) simple Lie superalgebras over algebraically closed fields of characteristic~${p=2}$. If $p\neq 2$, the ``superization'' $2 \longleftrightarrow 0$ may turn a finite-dimensional Lie algebra into an infinite-dimensional one or change the rate of its growth, but changes \eqref{step1} preserve growth in the cases considered, as we will see.

The ``desuperization'' we consider is the inverse of the ``superization'' procedure. (NB: both these procedures are completely different from their namesakes considered in \cite{BGLLS}.)

For affine Kac--Moody superalgebras, these ``desuperization''$ \longleftrightarrow$``superization'' procedures, first described by J.~van de Leur \cite{vdL}, are one-to-one in both directions for Lie (super)algebras whose Cartan matrices have no zeros on the main diagonal, except for the matrix $\left(\begin{smallmatrix} 2&-2\\-2&2\end{smallmatrix}\right)$ which obviously allows~2 non-equivalent ``superizations'' of 3 possible cases.
For the following hyperbolic Lie algebras, and almost affine Lie superalgebras, numbered as in \cite{CCLL} and in the next section, the ``desuperization''$ \longleftrightarrow$ ``superization'' correspondences are not one-to-one: each has at least~2 (the number in parentheses if $>2$) ``superizations'':
\begin{equation}\label{mult}
\begin{split}
& H3_{27}, \ H3_{87}, \ H3_{93} (3), \ H3_{98} (3), \ H3_{108} (3), \
H3_{113} (5), \ H3_{115} (3), \ H3_{117} (3), \\
& H3_{120} (3), \
H3_{123}, \ H4_{5} (3), \ H4_{16} (3), \ H4_{24}, \ H4_{44}, \ H4_{45}, \ H4_{46} (3), \ H5_{22} (3), \ H6_{9}.
\end{split}
\end{equation}

Observe that the properties of $\fosp(1|2n)$ resemble properties of the simple finite-dimensional Lie algebras, see \cite{DLZ}, where the mechanism of this resemblance found in earlier works is clarified; for the same reasons, we conjecture that the properties of almost affine Lie superalgebras we consider here are close to similar properties of hyperbolic Lie algebras.

\section[The pairs ``almost affine Lie superalgebra longleftrightarrow hyperbolic Lie algebra'' for symmetrizable Cartan matrices]{The pairs ``almost affine Lie superalgebra $\boldsymbol{\longleftrightarrow}$ hyperbolic\\ Lie algebra'' for symmetrizable Cartan matrices}\label{corresp}

In this section, we list all pairs ``almost affine Lie superalgebra $\longleftrightarrow$ hyperbolic Lie algebra''. In this list, the first column contains the name of an almost affine Lie superalgebra $\fg$, see \cite{CCLL}, the second one contains its Cartan matrix, the third one contains the name of the hyperbolic Lie algebra, see \cite{CCLL}, equal to $\fg_\ev$. The penultimate column contains the Cartan matrix of $\fg_\ev$.
The last column contains the permutation of rows/columns of the given $S$-type matrix of the first column after its row with 1 on the main diagonal is multiplied by 2. We mark the Lie algebras that admit several ``superizations''
by a~$\fbox{!!!}$. To sift through the list of Cartan matrices of hyperbolic Lie algebras and almost affine Lie superalgebras is possible by bare hands, but is rather boring and prone to inadvertent omissions, so we used a simple code in order to select all pairs suitable for the ``superization''$ \longleftrightarrow$ ``desuperization'' thus defined, see \cite{Lo}. We spotted a~typo in \cite{CCLL}: the matrix $S3_{35}$ is non-symmetrizable, should be designated $NS3_{35}$.

{\tiny
\[\arraycolsep=1.5pt
\begin{matrix}
S3_{4} & \smat{
 2 & -1 & -2 \\
 -1 & 1 & -1 \\
 -2 & -1 & 2 \\
} & H3_{4} & \smat{
 2 & -2 & -1 \\
 -2 & 2 & -1 \\
 -2 & -2 & 2 \\
} & \{1,3,2\} & \qquad S3_{5} & \smat{
 2 & -1 & -1 \\
 -1 & 1 & -1 \\
 -1 & -1 & 2 \\
} & H3_{5} & \smat{
 2 & -1 & -1 \\
 -1 & 2 & -1 \\
 -2 & -2 & 2 \\
} & \{1,3,2\} \\ \\
S3_{6} & \smat{
 2 & -1 & -1 \\
 -2 & 1 & -1 \\
 -2 & -1 & 2 \\
} & H3_{14} & \smat{
 2 & -1 & -1 \\
 -2 & 2 & -1 \\
 -4 & -2 & 2 \\
} & \{1,3,2\} & \qquad \fbox{!!!}~~S3_{19} & \smat{
 2 & -2 & -2 \\
 -1 & 1 & -1 \\
 -1 & -1 & 2 \\
} & H3_{27} & \smat{
 2 & -2 & -2 \\
 -1 & 2 & -1 \\
 -2 & -2 & 2 \\
} & \{1,3,2\} \\ \\
S3_{25} & \smat{
 2 & -1 & -2 \\
 -2 & 1 & -2 \\
 -2 & -1 & 2 \\
} & H3_{82} & \smat{
 2 & -2 & -1 \\
 -2 & 2 & -1 \\
 -4 & -4 & 2 \\
} & \{1,3,2\} & \qquad S3_{26} & \smat{
 2 & -1 & -1 \\
 -2 & 1 & -2 \\
 -1 & -1 & 2 \\
} & H3_{83} & \smat{
 2 & -1 & -1 \\
 -1 & 2 & -1 \\
 -4 & -4 & 2 \\
} & \{1,3,2\} \\
\\
\fbox{!!!}~~S3_{34} & \smat{
 2 & -2 & -4 \\
 -1 & 1 & -2 \\
 -1 & -1 & 2 \\
} & H3_{87} & \smat{
 2 & -4 & -2 \\
 -1 & 2 & -1 \\
 -2 & -4 & 2 \\
} & \{1,3,2\} & \qquad \fbox{!!!}~~S3_{38} & \smat{
 2 & -2 & -2 \\
 -1 & 1 & -1 \\
 -2 & -2 & 2 \\
} & H3_{93} & \smat{
 2 & -2 & -2 \\
 -2 & 2 & -2 \\
 -2 & -2 & 2 \\
} & \\ \\
S3_{39} & \smat{
 2 & -2 & -1 \\
 -1 & 1 & -1 \\
 -1 & -2 & 2 \\
} & H3_{94} & \smat{
 2 & -1 & -2 \\
 -1 & 2 & -2 \\
 -2 & -2 & 2 \\
} & \{1,3,2\} & \qquad \fbox{!!!}~~S3_{40} & \smat{
 2 & -1 & -1 \\
 -1 & 1 & -1 \\
 -1 & -1 & 1 \\
} & H3_{27} & \smat{
 2 & -2 & -2 \\
 -1 & 2 & -1 \\
 -2 & -2 & 2 \\
} & \{2,1,3\} \\
\\
\fbox{!!!}~~S3_{43} & \smat{
 2 & -1 & -1 \\
 -2 & 1 & -1 \\
 -2 & -1 & 1 \\
} & H3_{87} & \smat{
 2 & -4 & -2 \\
 -1 & 2 & -1 \\
 -2 & -4 & 2 \\
} & \{2,1,3\} & \qquad \fbox{!!!}~~S3_{45} & \smat{
 2 & -2 & -2 \\
 -1 & 1 & -1 \\
 -1 & -1 & 1 \\
} & H3_{93} & \smat{
 2 & -2 & -2 \\
 -2 & 2 & -2 \\
 -2 & -2 & 2 \\
} & \\ \\
\fbox{!!!}~~S3_{46} & \smat{
 1 & -1 & -1 \\
 -1 & 1 & -1 \\
 -1 & -1 & 1 \\
} & H3_{93} & \smat{
 2 & -2 & -2 \\
 -2 & 2 & -2 \\
 -2 & -2 & 2 \\
} & & \qquad S3_{47} & \smat{
 2 & -1 & 0 \\
 -2 & 1 & -1 \\
 0 & -1 & 2 \\
} & H3_{97} & \smat{
 2 & -1 & 0 \\
 -4 & 2 & -2 \\
 0 & -1 & 2 \\
} & \\ \\
\fbox{!!!}~~S3_{48} & \smat{
 2 & -2 & 0 \\
 -1 & 1 & -1 \\
 0 & -1 & 2 \\
} & H3_{98} & \smat{
 2 & -2 & 0 \\
 -2 & 2 & -2 \\
 0 & -1 & 2 \\
} & & \qquad S3_{49} & \smat{
 2 & -1 & 0 \\
 -2 & 1 & -2 \\
 0 & -1 & 2 \\
} & H3_{107} & \smat{
 2 & -1 & 0 \\
 -4 & 2 & -4 \\
 0 & -1 & 2 \\
} & \\
\\
\fbox{!!!}~~S3_{50} & \smat{
 2 & -2 & 0 \\
 -1 & 1 & -2 \\
 0 & -1 & 2 \\
} & H3_{108} & \smat{
 2 & -2 & 0 \\
 -2 & 2 & -4 \\
 0 & -1 & 2 \\
} & & \qquad \fbox{!!!}~~S3_{51} & \smat{
 2 & -2 & 0 \\
 -1 & 1 & -1 \\
 0 & -2 & 2 \\
} & H3_{113} & \smat{
 2 & -2 & 0 \\
 -2 & 2 & -2 \\
 0 & -2 & 2 \\
} & \\
\\
S3_{52} & \smat{
 2 & -1 & 0 \\
 -3 & 2 & -1 \\
 0 & -1 & 1 \\
} & H3_{104} & \smat{
 2 & -2 & 0 \\
 -1 & 2 & -3 \\
 0 & -1 & 2 \\
} & \{3,2,1\} & \qquad S3_{53} & \smat{
 2 & -1 & 0 \\
 -4 & 2 & -1 \\
 0 & -1 & 1 \\
} & H3_{110} & \smat{
 2 & -2 & 0 \\
 -1 & 2 & -4 \\
 0 & -1 & 2 \\
} & \{3,2,1\} \\
\end{matrix}
\]
\[\arraycolsep=1.5pt
\begin{matrix}
\fbox{!!!}~~S3_{54} & \smat{
 2 & -2 & 0 \\
 -2 & 2 & -1 \\
 0 & -1 & 1 \\
} & H3_{115} & \smat{
 2 & -2 & 0 \\
 -1 & 2 & -2 \\
 0 & -2 & 2 \\
} & \{3,2,1\} & \qquad S3_{55} & \smat{
 2 & -3 & 0 \\
 -1 & 2 & -1 \\
 0 & -1 & 1 \\
} & H3_{119} & \smat{
 2 & -3 & 0 \\
 -1 & 2 & -1 \\
 0 & -2 & 2 \\
} & \\ \\
\fbox{!!!}~~S3_{56} & \smat{
 2 & -4 & 0 \\
 -1 & 2 & -1 \\
 0 & -1 & 1 \\
} & H3_{120} & \smat{
 2 & -4 & 0 \\
 -1 & 2 & -1 \\
 0 & -2 & 2 \\
} & & \qquad S3_{57} & \smat{
 2 & -1 & 0 \\
 -2 & 2 & -1 \\
 0 & -2 & 1 \\
} & H3_{100} & \smat{
 2 & -4 & 0 \\
 -1 & 2 & -2 \\
 0 & -1 & 2 \\
} & \{3,2,1\}
\\
S3_{58} & \smat{
 2 & -1 & 0 \\
 -3 & 2 & -1 \\
 0 & -2 & 1 \\
} & H3_{106} & \smat{
 2 & -4 & 0 \\
 -1 & 2 & -3 \\
 0 & -1 & 2 \\
} & \{3,2,1\} & \qquad S3_{59} & \smat{
 2 & -1 & 0 \\
 -4 & 2 & -1 \\
 0 & -2 & 1 \\
} & H3_{112} & \smat{
 2 & -4 & 0 \\
 -1 & 2 & -4 \\
 0 & -1 & 2 \\
} & \{3,2,1\} \\ \\
\fbox{!!!}~~S3_{60} & \smat{
 2 & -2 & 0 \\
 -2 & 2 & -1 \\
 0 & -2 & 1 \\
} & H3_{117} & \smat{
 2 & -4 & 0 \\
 -1 & 2 & -2 \\
 0 & -2 & 2 \\
} & \{3,2,1\} & \qquad S3_{61} & \smat{
 2 & -1 & 0 \\
 -1 & 2 & -1 \\
 0 & -2 & 1 \\
} & H3_{118} & \smat{
 2 & -4 & 0 \\
 -1 & 2 & -1 \\
 0 & -1 & 2 \\
} & \{3,2,1\} \\ \\
\fbox{!!!}~~S3_{62} & \smat{
 2 & -2 & 0 \\
 -1 & 2 & -1 \\
 0 & -2 & 1 \\
} & H3_{120} & \smat{
 2 & -4 & 0 \\
 -1 & 2 & -1 \\
 0 & -2 & 2 \\
} & \{3,2,1\} & \qquad S3_{63} & \smat{
 2 & -3 & 0 \\
 -1 & 2 & -1 \\
 0 & -2 & 1 \\
} & H3_{122} & \smat{
 2 & -4 & 0 \\
 -1 & 2 & -1 \\
 0 & -3 & 2 \\
} & \{3,2,1\} \\ \\
\fbox{!!!}~~S3_{64} & \smat{
 2 & -4 & 0 \\
 -1 & 2 & -1 \\
 0 & -2 & 1 \\
} & H3_{123} & \smat{
 2 & -4 & 0 \\
 -1 & 2 & -1 \\
 0 & -4 & 2 \\
} & & \qquad \fbox{!!!}~~S3_{65} & \smat{
 2 & -1 & 0 \\
 -2 & 2 & -2 \\
 0 & -1 & 1 \\
} & H3_{98} & \smat{
 2 & -2 & 0 \\
 -2 & 2 & -2 \\
 0 & -1 & 2 \\
} & \{3,2,1\} \\ \\
S3_{66} & \smat{
 2 & -1 & 0 \\
 -3 & 2 & -2 \\
 0 & -1 & 1 \\
} & H3_{103} & \smat{
 2 & -2 & 0 \\
 -2 & 2 & -3 \\
 0 & -1 & 2 \\
} & \{3,2,1\} & \qquad \fbox{!!!}~~S3_{67} & \smat{
 2 & -1 & 0 \\
 -4 & 2 & -2 \\
 0 & -1 & 1 \\
} & H3_{108} & \smat{
 2 & -2 & 0 \\
 -2 & 2 & -4 \\
 0 & -1 & 2 \\
} & \{3,2,1\} \\ \\
\fbox{!!!}~~S3_{68} & \smat{
 2 & -2 & 0 \\
 -2 & 2 & -2 \\
 0 & -1 & 1 \\
} & H3_{113} & \smat{
 2 & -2 & 0 \\
 -2 & 2 & -2 \\
 0 & -2 & 2 \\
} & & \qquad S3_{69} & \smat{
 2 & -1 & 0 \\
 -1 & 2 & -2 \\
 0 & -1 & 1 \\
} & H3_{114} & \smat{
 2 & -1 & 0 \\
 -1 & 2 & -2 \\
 0 & -2 & 2 \\
} & \\ \\
\fbox{!!!}~~S3_{70} & \smat{
 2 & -2 & 0 \\
 -1 & 2 & -2 \\
 0 & -1 & 1 \\
} & H3_{115} & \smat{
 2 & -2 & 0 \\
 -1 & 2 & -2 \\
 0 & -2 & 2 \\
} & & \qquad S3_{71} & \smat{
 2 & -3 & 0 \\
 -1 & 2 & -2 \\
 0 & -1 & 1 \\
} & H3_{116} & \smat{
 2 & -3 & 0 \\
 -1 & 2 & -2 \\
 0 & -2 & 2 \\
} & \\ \\
\fbox{!!!}~~S3_{72} & \smat{
 2 & -4 & 0 \\
 -1 & 2 & -2 \\
 0 & -1 & 1 \\
} & H3_{117} & \smat{
 2 & -4 & 0 \\
 -1 & 2 & -2 \\
 0 & -2 & 2 \\
} & & \qquad \fbox{!!!}~~S3_{73} & \smat{
 1 & -2 & 0 \\
 -1 & 2 & -1 \\
 0 & -1 & 1 \\
} & H3_{120} & \smat{
 2 & -4 & 0 \\
 -1 & 2 & -1 \\
 0 & -2 & 2 \\
} & \\ \\
\fbox{!!!}~~S3_{74} & \smat{
 1 & -1 & 0 \\
 -2 & 2 & -1 \\
 0 & -1 & 1 \\
} & H3_{115} & \smat{
 2 & -2 & 0 \\
 -1 & 2 & -2 \\
 0 & -2 & 2 \\
} & \{3,2,1\} & \qquad \fbox{!!!}~~S3_{75} & \smat{
 1 & -2 & 0 \\
 -1 & 2 & -1 \\
 0 & -2 & 1 \\
} & H3_{123} & \smat{
 2 & -4 & 0 \\
 -1 & 2 & -1 \\
 0 & -4 & 2 \\
} & \\ \\
\fbox{!!!}~~S3_{76} & \smat{
 1 & -1 & 0 \\
 -2 & 2 & -1 \\
 0 & -2 & 1 \\
} & H3_{117} & \smat{
 2 & -4 & 0 \\
 -1 & 2 & -2 \\
 0 & -2 & 2 \\
} & \{3,2,1\} & \qquad \fbox{!!!}~~S3_{77} & \smat{
 1 & -1 & 0 \\
 -2 & 2 & -2 \\
 0 & -1 & 1 \\
} & H3_{113} & \smat{
 2 & -2 & 0 \\
 -2 & 2 & -2 \\
 0 & -2 & 2 \\
} & \\ \\
\fbox{!!!}~~S3_{78} & \smat{
 2 & -1 & 0 \\
 -1 & 1 & -1 \\
 0 & -1 & 1 \\
} & H3_{98} & \smat{
 2 & -2 & 0 \\
 -2 & 2 & -2 \\
 0 & -1 & 2 \\
} & \{3,2,1\} & \qquad \fbox{!!!}~~S3_{79} & \smat{
 2 & -1 & 0 \\
 -2 & 1 & -1 \\
 0 & -1 & 1 \\
} & H3_{108} & \smat{
 2 & -2 & 0 \\
 -2 & 2 & -4 \\
 0 & -1 & 2 \\
} & \{3,2,1\} \\ \\
\fbox{!!!}~~S3_{80} & \smat{
 2 & -2 & 0 \\
 -1 & 1 & -1 \\
 0 & -1 & 1 \\
} & H3_{113} & \smat{
 2 & -2 & 0 \\
 -2 & 2 & -2 \\
 0 & -2 & 2 \\
} & & \qquad \fbox{!!!}~~S3_{81} & \smat{
 1 & -1 & 0 \\
 -1 & 1 & -1 \\
 0 & -1 & 1 \\
} & H3_{113} & \smat{
 2 & -2 & 0 \\
 -2 & 2 & -2 \\
 0 & -2 & 2 \\
} &
\\ \\
S4_{1} & \smat{
 2 & -1 & 0 & 0 \\
 -1 & 1 & -1 & 0 \\
 0 & -1 & 2 & -1 \\
 0 & 0 & -1 & 2 \\
} & H4_{3} & \smat{
 2 & -1 & 0 & 0 \\
 -1 & 2 & -1 & 0 \\
 0 & -2 & 2 & -2 \\
 0 & 0 & -1 & 2 \\
} & \{4,3,2,1\} & \qquad S4_{2} & \smat{
 2 & -1 & 0 & 0 \\
 -1 & 1 & -1 & 0 \\
 0 & -1 & 2 & -2 \\
 0 & 0 & -1 & 2 \\
} & H4_{1} & \smat{
 2 & -1 & 0 & 0 \\
 -2 & 2 & -1 & 0 \\
 0 & -2 & 2 & -2 \\
 0 & 0 & -1 & 2 \\
} & \{4,3,2,1\} \\ \\
\fbox{!!!}~~S4_{3} & \smat{
 2 & -1 & 0 & 0 \\
 -1 & 1 & -1 & 0 \\
 0 & -1 & 2 & -1 \\
 0 & 0 & -2 & 2 \\
} & H4_{5} & \smat{
 2 & -2 & 0 & 0 \\
 -1 & 2 & -1 & 0 \\
 0 & -2 & 2 & -2 \\
 0 & 0 & -1 & 2 \\
} & \{4,3,2,1\} & \qquad S4_{4} & \smat{
 2 & -1 & 0 & 0 \\
 -3 & 2 & -1 & 0 \\
 0 & -1 & 2 & -1 \\
 0 & 0 & -1 & 1 \\
} & H4_{10} & \smat{
 2 & -2 & 0 & 0 \\
 -1 & 2 & -1 & 0 \\
 0 & -1 & 2 & -3 \\
 0 & 0 & -1 & 2 \\
} & \{4,3,2,1\} \\ \\
S4_{5} & \smat{
 2 & -3 & 0 & 0 \\
 -1 & 2 & -1 & 0 \\
 0 & -1 & 2 & -1 \\
 0 & 0 & -1 & 1 \\
} & H4_{17} & \smat{
 2 & -3 & 0 & 0 \\
 -1 & 2 & -1 & 0 \\
 0 & -1 & 2 & -1 \\
 0 & 0 & -2 & 2 \\
} & & \qquad S4_{6} & \smat{
 2 & -1 & 0 & 0 \\
 -2 & 2 & -1 & 0 \\
 0 & -2 & 2 & -1 \\
 0 & 0 & -1 & 1 \\
} & H4_{6} & \smat{
 2 & -2 & 0 & 0 \\
 -1 & 2 & -2 & 0 \\
 0 & -1 & 2 & -2 \\
 0 & 0 & -1 & 2 \\
} & \{4,3,2,1\} \\
\\
S4_{7} & \smat{
 2 & -1 & 0 & 0 \\
 -1 & 2 & -1 & 0 \\
 0 & -2 & 2 & -1 \\
 0 & 0 & -1 & 1 \\
} & H4_{13} & \smat{
 2 & -2 & 0 & 0 \\
 -1 & 2 & -2 & 0 \\
 0 & -1 & 2 & -1 \\
 0 & 0 & -1 & 2 \\
} & \{4,3,2,1\} & \qquad \fbox{!!!}~~S4_{8} & \smat{
 2 & -2 & 0 & 0 \\
 -1 & 2 & -1 & 0 \\
 0 & -2 & 2 & -1 \\
 0 & 0 & -1 & 1 \\
} & H4_{16} & \smat{
 2 & -2 & 0 & 0 \\
 -1 & 2 & -2 & 0 \\
 0 & -1 & 2 & -1 \\
 0 & 0 & -2 & 2 \\
} & \{4,3,2,1\} \\ \\
\fbox{!!!}~~S4_{9} & \smat{
 2 & -1 & 0 & 0 \\
 -2 & 2 & -2 & 0 \\
 0 & -1 & 2 & -1 \\
 0 & 0 & -1 & 1 \\
} & H4_{5} & \smat{
 2 & -2 & 0 & 0 \\
 -1 & 2 & -1 & 0 \\
 0 & -2 & 2 & -2 \\
 0 & 0 & -1 & 2 \\
} & \{4,3,2,1\} & \qquad S4_{10} & \smat{
 2 & -1 & 0 & 0 \\
 -1 & 2 & -2 & 0 \\
 0 & -1 & 2 & -1 \\
 0 & 0 & -1 & 1 \\
} & H4_{14} & \smat{
 2 & -2 & 0 & 0 \\
 -1 & 2 & -1 & 0 \\
 0 & -2 & 2 & -1 \\
 0 & 0 & -1 & 2 \\
} & \{4,3,2,1\} \\
\\
\fbox{!!!}~~S4_{11} & \smat{
 2 & -2 & 0 & 0 \\
 -1 & 2 & -2 & 0 \\
 0 & -1 & 2 & -1 \\
 0 & 0 & -1 & 1 \\
} & H4_{16} & \smat{
 2 & -2 & 0 & 0 \\
 -1 & 2 & -2 & 0 \\
 0 & -1 & 2 & -1 \\
 0 & 0 & -2 & 2 \\
} & & \qquad \fbox{!!!}~~S4_{12} & \smat{
 1 & -1 & 0 & 0 \\
 -1 & 2 & -1 & 0 \\
 0 & -2 & 2 & -1 \\
 0 & 0 & -1 & 1 \\
} & H4_{16} & \smat{
 2 & -2 & 0 & 0 \\
 -1 & 2 & -2 & 0 \\
 0 & -1 & 2 & -1 \\
 0 & 0 & -2 & 2 \\
} & \{4,3,2,1\} \\ \\
\fbox{!!!}~~S4_{13} & \smat{
 2 & -1 & 0 & 0 \\
 -1 & 1 & -1 & 0 \\
 0 & -1 & 2 & -1 \\
 0 & 0 & -1 & 1 \\
} & H4_{5} & \smat{
 2 & -2 & 0 & 0 \\
 -1 & 2 & -1 & 0 \\
 0 & -2 & 2 & -2 \\
 0 & 0 & -1 & 2 \\
} & \{4,3,2,1\} & \qquad S4_{14} & \smat{
 2 & -1 & 0 & -1 \\
 -1 & 1 & -1 & 0 \\
 0 & -1 & 2 & -1 \\
 -1 & 0 & -1 & 2 \\
} & H4_{19} & \smat{
 2 & -1 & 0 & -1 \\
 -1 & 2 & -1 & 0 \\
 0 & -2 & 2 & -2 \\
 -1 & 0 & -1 & 2 \\
} & \{2,3,4,1\} \\
\\
S4_{17} & \smat{
 2 & -1 & 0 & -2 \\
 -1 & 1 & -1 & 0 \\
 0 & -1 & 2 & -2 \\
 -1 & 0 & -1 & 2 \\
} & H4_{22} & \smat{
 2 & -1 & 0 & -1 \\
 -2 & 2 & -1 & 0 \\
 0 & -2 & 2 & -2 \\
 -2 & 0 & -1 & 2 \\
} & \{2,3,4,1\} & \qquad \fbox{!!!}~~S4_{19} & \smat{
 2 & -1 & 0 & -1 \\
 -1 & 1 & -1 & 0 \\
 0 & -1 & 2 & -1 \\
 -2 & 0 & -2 & 2 \\
} & H4_{24} & \smat{
 2 & -2 & 0 & -2 \\
 -1 & 2 & -1 & 0 \\
 0 & -2 & 2 & -2 \\
 -1 & 0 & -1 & 2 \\
} & \{2,1,4,3\} \\ \\
\fbox{!!!}~~S4_{20} & \smat{
 2 & -1 & 0 & -1 \\
 -1 & 1 & -1 & 0 \\
 0 & -1 & 2 & -1 \\
 -1 & 0 & -1 & 1 \\
} & H4_{24} & \smat{
 2 & -2 & 0 & -2 \\
 -1 & 2 & -1 & 0 \\
 0 & -2 & 2 & -2 \\
 -1 & 0 & -1 & 2 \\
} & \{2,1,4,3\} & \qquad S4_{21} & \smat{
 2 & -1 & 0 & 0 \\
 -1 & 1 & -1 & -1 \\
 0 & -1 & 2 & 0 \\
 0 & -1 & 0 & 2 \\
} & H4_{38} & \smat{
 2 & -1 & 0 & 0 \\
 -2 & 2 & -2 & -2 \\
 0 & -1 & 2 & 0 \\
 0 & -1 & 0 & 2 \\
} & \\
\\
S4_{22} & \smat{
 2 & -1 & 0 & 0 \\
 -2 & 2 & -1 & -1 \\
 0 & -1 & 1 & 0 \\
 0 & -1 & 0 & 2 \\
} & H4_{43} & \smat{
 2 & -1 & 0 & 0 \\
 -2 & 2 & -1 & -1 \\
 0 & -1 & 2 & 0 \\
 0 & -2 & 0 & 2 \\
} & \{1,2,4,3\} & \qquad \fbox{!!!}~~S4_{23} & \smat{
 2 & -2 & 0 & 0 \\
 -1 & 2 & -1 & -1 \\
 0 & -1 & 1 & 0 \\
 0 & -1 & 0 & 2 \\
} & H4_{44} & \smat{
 2 & -2 & 0 & 0 \\
 -1 & 2 & -1 & -1 \\
 0 & -1 & 2 & 0 \\
 0 & -2 & 0 & 2 \\
} & \{1,2,4,3\} \\ \\
S4_{24} & \smat{
 2 & -1 & 0 & 0 \\
 -2 & 2 & -1 & -2 \\
 0 & -1 & 1 & 0 \\
 0 & -1 & 0 & 2 \\
} & H4_{40} & \smat{
 2 & -2 & 0 & 0 \\
 -1 & 2 & -2 & -2 \\
 0 & -1 & 2 & 0 \\
 0 & -1 & 0 & 2 \\
} & \{3,2,1,4\} & \qquad \fbox{!!!}~~S4_{25} & \smat{
 2 & -2 & 0 & 0 \\
 -1 & 2 & -1 & -2 \\
 0 & -1 & 1 & 0 \\
 0 & -1 & 0 & 2 \\
} & H4_{45} & \smat{
 2 & -2 & 0 & 0 \\
 -1 & 2 & -2 & -1 \\
 0 & -1 & 2 & 0 \\
 0 & -2 & 0 & 2 \\
} & \{1,2,4,3\} \\
\\
\fbox{!!!}~~S4_{26} & \smat{
 2 & -2 & 0 & 0 \\
 -1 & 2 & -1 & -1 \\
 0 & -1 & 1 & 0 \\
 0 & -2 & 0 & 2 \\
} & H4_{46} & \smat{
 2 & -2 & 0 & 0 \\
 -1 & 2 & -1 & -1 \\
 0 & -2 & 2 & 0 \\
 0 & -2 & 0 & 2 \\
} & & \qquad \fbox{!!!}~~S4_{27} & \smat{
 2 & -1 & 0 & 0 \\
 -1 & 2 & -1 & -1 \\
 0 & -1 & 1 & 0 \\
 0 & -1 & 0 & 1 \\
} & H4_{44} & \smat{
 2 & -2 & 0 & 0 \\
 -1 & 2 & -1 & -1 \\
 0 & -1 & 2 & 0 \\
 0 & -2 & 0 & 2 \\
} & \{3,2,1,4\} \\
\\
\fbox{!!!}~~S4_{28} & \smat{
 2 & -1 & 0 & 0 \\
 -2 & 2 & -1 & -1 \\
 0 & -1 & 1 & 0 \\
 0 & -1 & 0 & 1 \\
} & H4_{45} & \smat{
 2 & -2 & 0 & 0 \\
 -1 & 2 & -2 & -1 \\
 0 & -1 & 2 & 0 \\
 0 & -2 & 0 & 2 \\
} & \{3,2,1,4\} & \qquad \fbox{!!!}~~S4_{29} & \smat{
 2 & -2 & 0 & 0 \\
 -1 & 2 & -1 & -1 \\
 0 & -1 & 1 & 0 \\
 0 & -1 & 0 & 1 \\
} & H4_{46} & \smat{
 2 & -2 & 0 & 0 \\
 -1 & 2 & -1 & -1 \\
 0 & -2 & 2 & 0 \\
 0 & -2 & 0 & 2 \\
} & \\
\\
\fbox{!!!}~~S4_{30} & \smat{
 1 & -1 & 0 & 0 \\
 -1 & 2 & -1 & -1 \\
 0 & -1 & 1 & 0 \\
 0 & -1 & 0 & 1 \\
} & H4_{46} & \smat{
 2 & -2 & 0 & 0 \\
 -1 & 2 & -1 & -1 \\
 0 & -2 & 2 & 0 \\
 0 & -2 & 0 & 2 \\
} & & \qquad S4_{31} & \smat{
 1 & -1 & 0 & 0 \\
 -1 & 2 & -1 & -1 \\
 0 & -1 & 2 & -1 \\
 0 & -1 & -1 & 2 \\
} & H4_{52} & \smat{
 2 & -2 & 0 & 0 \\
 -1 & 2 & -1 & -1 \\
 0 & -1 & 2 & -1 \\
 0 & -1 & -1 & 2 \\
} &
\end{matrix}
\]
\[\arraycolsep=1.5pt
\begin{matrix}
S5_{1} & \smat{
 2 & -1 & 0 & 0 & 0 \\
 -1 & 2 & -1 & 0 & 0 \\
 0 & -2 & 2 & -1 & 0 \\
 0 & 0 & -1 & 2 & -1 \\
 0 & 0 & 0 & -1 & 1 \\
} & H5_{4} & \smat{
 2 & -1 & 0 & 0 & 0 \\
 -1 & 2 & -1 & 0 & 0 \\
 0 & -2 & 2 & -1 & 0 \\
 0 & 0 & -1 & 2 & -1 \\
 0 & 0 & 0 & -2 & 2 \\
} & \\ \\
S5_{2} & \smat{
 2 & -1 & 0 & 0 & 0 \\
 -1 & 2 & -2 & 0 & 0 \\
 0 & -1 & 2 & -1 & 0 \\
 0 & 0 & -1 & 2 & -1 \\
 0 & 0 & 0 & -1 & 1 \\
} & H5_{3} & \smat{
 2 & -1 & 0 & 0 & 0 \\
 -1 & 2 & -2 & 0 & 0 \\
 0 & -1 & 2 & -1 & 0 \\
 0 & 0 & -1 & 2 & -1 \\
 0 & 0 & 0 & -2 & 2 \\
} & \\ \\
S5_{3} & \smat{
 2 & -1 & 0 & 0 & 0 \\
 -1 & 2 & -1 & -1 & -1 \\
 0 & -1 & 1 & 0 & 0 \\
 0 & -1 & 0 & 2 & 0 \\
 0 & -1 & 0 & 0 & 2 \\
} & H5_{10} & \smat{
 2 & -2 & 0 & 0 & 0 \\
 -1 & 2 & -1 & -1 & -1 \\
 0 & -1 & 2 & 0 & 0 \\
 0 & -1 & 0 & 2 & 0 \\
 0 & -1 & 0 & 0 & 2 \\
} & \{3,2,1,4,5\} \\ \\
S5_{4} & \smat{
 2 & -1 & 0 & 0 & 0 \\
 -2 & 2 & -1 & -1 & 0 \\
 0 & -1 & 2 & 0 & 0 \\
 0 & -1 & 0 & 2 & -1 \\
 0 & 0 & 0 & -1 & 1 \\
} & H5_{19} & \smat{
 2 & -1 & 0 & 0 & 0 \\
 -1 & 2 & -2 & -1 & 0 \\
 0 & -1 & 2 & 0 & 0 \\
 0 & -1 & 0 & 2 & -1 \\
 0 & 0 & 0 & -2 & 2 \\
} & \{3,2,1,4,5\} \\ \\
\fbox{!!!}~~S5_{5} & \smat{
 2 & -2 & 0 & 0 & 0 \\
 -1 & 2 & -1 & -1 & 0 \\
 0 & -1 & 2 & 0 & 0 \\
 0 & -1 & 0 & 2 & -1 \\
 0 & 0 & 0 & -1 & 1 \\
} & H5_{22} & \smat{
 2 & -1 & 0 & 0 & 0 \\
 -1 & 2 & -1 & -1 & 0 \\
 0 & -2 & 2 & 0 & 0 \\
 0 & -1 & 0 & 2 & -1 \\
 0 & 0 & 0 & -2 & 2 \\
} & \{3,2,1,4,5\} \\ \\
S5_{6} & \smat{
 2 & -1 & 0 & 0 & 0 \\
 -1 & 2 & -1 & -1 & 0 \\
 0 & -1 & 1 & 0 & 0 \\
 0 & -1 & 0 & 2 & -1 \\
 0 & 0 & 0 & -1 & 2 \\
} & H5_{21} & \smat{
 2 & -1 & 0 & 0 & 0 \\
 -1 & 2 & -1 & -1 & 0 \\
 0 & -2 & 2 & 0 & 0 \\
 0 & -1 & 0 & 2 & -1 \\
 0 & 0 & 0 & -1 & 2 \\
} & \\ \\
S5_{7} & \smat{
 2 & -1 & 0 & 0 & 0 \\
 -1 & 2 & -1 & -1 & 0 \\
 0 & -1 & 1 & 0 & 0 \\
 0 & -1 & 0 & 2 & -2 \\
 0 & 0 & 0 & -1 & 2 \\
} & H5_{20} & \smat{
 2 & -1 & 0 & 0 & 0 \\
 -1 & 2 & -1 & -1 & 0 \\
 0 & -2 & 2 & 0 & 0 \\
 0 & -1 & 0 & 2 & -2 \\
 0 & 0 & 0 & -1 & 2 \\
} & \\ \\
\fbox{!!!}~~S5_{8} & \smat{
 2 & -1 & 0 & 0 & 0 \\
 -1 & 2 & -1 & -1 & 0 \\
 0 & -1 & 1 & 0 & 0 \\
 0 & -1 & 0 & 2 & -1 \\
 0 & 0 & 0 & -2 & 2 \\
} & H5_{22} & \smat{
 2 & -1 & 0 & 0 & 0 \\
 -1 & 2 & -1 & -1 & 0 \\
 0 & -2 & 2 & 0 & 0 \\
 0 & -1 & 0 & 2 & -1 \\
 0 & 0 & 0 & -2 & 2 \\
} & \\ \\
\fbox{!!!}~~S5_{9} & \smat{
 2 & -1 & 0 & 0 & 0 \\
 -1 & 2 & -1 & -1 & 0 \\
 0 & -1 & 1 & 0 & 0 \\
 0 & -1 & 0 & 2 & -1 \\
 0 & 0 & 0 & -1 & 1 \\
} & H5_{22} & \smat{
 2 & -1 & 0 & 0 & 0 \\
 -1 & 2 & -1 & -1 & 0 \\
 0 & -2 & 2 & 0 & 0 \\
 0 & -1 & 0 & 2 & -1 \\
 0 & 0 & 0 & -2 & 2 \\
} & \\ \\
S5_{10} & \smat{
 1 & -1 & 0 & 0 & 0 \\
 -1 & 2 & -1 & 0 & -1 \\
 0 & -1 & 2 & -1 & 0 \\
 0 & 0 & -1 & 2 & -1 \\
 0 & -1 & 0 & -1 & 2 \\
} & H5_{14} & \smat{
 2 & -2 & 0 & 0 & 0 \\
 -1 & 2 & -1 & 0 & -1 \\
 0 & -1 & 2 & -1 & 0 \\
 0 & 0 & -1 & 2 & -1 \\
 0 & -1 & 0 & -1 & 2 \\
} &
\end{matrix}
\]
\[\arraycolsep=1.5pt
\begin{matrix}
S6_{1} & \smat{
 2 & -1 & 0 & 0 & 0 & 0 \\
 -1 & 2 & -1 & -1 & -1 & 0 \\
 0 & -1 & 2 & 0 & 0 & 0 \\
 0 & -1 & 0 & 2 & 0 & 0 \\
 0 & -1 & 0 & 0 & 2 & -1 \\
 0 & 0 & 0 & 0 & -1 & 1 \\
} & H6_{21} & \smat{
 2 & -1 & 0 & 0 & 0 & 0 \\
 -1 & 2 & -1 & -1 & -1 & 0 \\
 0 & -1 & 2 & 0 & 0 & 0 \\
 0 & -1 & 0 & 2 & 0 & 0 \\
 0 & -1 & 0 & 0 & 2 & -1 \\
 0 & 0 & 0 & 0 & -2 & 2 \\
} & \\ \\
S6_{2} & \smat{
 2 & -1 & 0 & 0 & 0 & 0 \\
 -1 & 2 & -1 & 0 & 0 & 0 \\
 0 & -1 & 2 & -1 & -1 & 0 \\
 0 & 0 & -1 & 2 & 0 & 0 \\
 0 & 0 & -1 & 0 & 2 & -1 \\
 0 & 0 & 0 & 0 & -1 & 1 \\
} & H6_{7} & \smat{
 2 & -2 & 0 & 0 & 0 & 0 \\
 -1 & 2 & -1 & 0 & 0 & 0 \\
 0 & -1 & 2 & -1 & -1 & 0 \\
 0 & 0 & -1 & 2 & 0 & 0 \\
 0 & 0 & -1 & 0 & 2 & -1 \\
 0 & 0 & 0 & 0 & -1 & 2 \\
} & \{6,5,3,4,2,1\} \\ \\
S6_{3} & \smat{
 2 & -1 & 0 & 0 & 0 & 0 \\
 -2 & 2 & -1 & 0 & 0 & 0 \\
 0 & -1 & 2 & -1 & -1 & 0 \\
 0 & 0 & -1 & 2 & 0 & 0 \\
 0 & 0 & -1 & 0 & 2 & -1 \\
 0 & 0 & 0 & 0 & -1 & 1 \\
} & H6_{10} & \smat{
 2 & -1 & 0 & 0 & 0 & 0 \\
 -2 & 2 & -1 & 0 & 0 & 0 \\
 0 & -1 & 2 & -1 & -1 & 0 \\
 0 & 0 & -1 & 2 & 0 & 0 \\
 0 & 0 & -1 & 0 & 2 & -1 \\
 0 & 0 & 0 & 0 & -2 & 2 \\
} & \\ \\
\fbox{!!!}~~S6_{4} & \smat{
 2 & -2 & 0 & 0 & 0 & 0 \\
 -1 & 2 & -1 & 0 & 0 & 0 \\
 0 & -1 & 2 & -1 & -1 & 0 \\
 0 & 0 & -1 & 2 & 0 & 0 \\
 0 & 0 & -1 & 0 & 2 & -1 \\
 0 & 0 & 0 & 0 & -1 & 1 \\
} & H6_{9} & \smat{
 2 & -2 & 0 & 0 & 0 & 0 \\
 -1 & 2 & -1 & 0 & 0 & 0 \\
 0 & -1 & 2 & -1 & -1 & 0 \\
 0 & 0 & -1 & 2 & 0 & 0 \\
 0 & 0 & -1 & 0 & 2 & -1 \\
 0 & 0 & 0 & 0 & -2 & 2 \\
} & \\ \\
\fbox{!!!}~~S6_{5} & \smat{
 1 & -1 & 0 & 0 & 0 & 0 \\
 -1 & 2 & -1 & 0 & 0 & 0 \\
 0 & -1 & 2 & -1 & -1 & 0 \\
 0 & 0 & -1 & 2 & 0 & 0 \\
 0 & 0 & -1 & 0 & 2 & -1 \\
 0 & 0 & 0 & 0 & -1 & 1 \\
} & H6_{9} & \smat{
 2 & -2 & 0 & 0 & 0 & 0 \\
 -1 & 2 & -1 & 0 & 0 & 0 \\
 0 & -1 & 2 & -1 & -1 & 0 \\
 0 & 0 & -1 & 2 & 0 & 0 \\
 0 & 0 & -1 & 0 & 2 & -1 \\
 0 & 0 & 0 & 0 & -2 & 2 \\
} & \\ \\
S6_{6} & \smat{
 2 & -1 & 0 & 0 & 0 & 0 \\
 -1 & 2 & -1 & 0 & 0 & 0 \\
 0 & -2 & 2 & -1 & 0 & 0 \\
 0 & 0 & -1 & 2 & -1 & 0 \\
 0 & 0 & 0 & -1 & 2 & -1 \\
 0 & 0 & 0 & 0 & -1 & 1 \\
} & H6_{18} & \smat{
 2 & -1 & 0 & 0 & 0 & 0 \\
 -1 & 2 & -1 & 0 & 0 & 0 \\
 0 & -2 & 2 & -1 & 0 & 0 \\
 0 & 0 & -1 & 2 & -1 & 0 \\
 0 & 0 & 0 & -1 & 2 & -1 \\
 0 & 0 & 0 & 0 & -2 & 2 \\
} & \\ \\
S6_{7} & \smat{
 2 & -1 & 0 & 0 & 0 & 0 \\
 -1 & 2 & -2 & 0 & 0 & 0 \\
 0 & -1 & 2 & -1 & 0 & 0 \\
 0 & 0 & -1 & 2 & -1 & 0 \\
 0 & 0 & 0 & -1 & 2 & -1 \\
 0 & 0 & 0 & 0 & -1 & 1 \\
} & H6_{17} & \smat{
 2 & -1 & 0 & 0 & 0 & 0 \\
 -1 & 2 & -2 & 0 & 0 & 0 \\
 0 & -1 & 2 & -1 & 0 & 0 \\
 0 & 0 & -1 & 2 & -1 & 0 \\
 0 & 0 & 0 & -1 & 2 & -1 \\
 0 & 0 & 0 & 0 & -2 & 2 \\
} &
\end{matrix}
\]
\[\arraycolsep=1.5pt
\begin{matrix}
S7_{1} & \smat{
 2 & -1 & 0 & 0 & 0 & 0 & 0 \\
 -1 & 2 & -1 & 0 & 0 & 0 & 0 \\
 0 & -1 & 2 & -1 & -1 & 0 & 0 \\
 0 & 0 & -1 & 2 & 0 & 0 & 0 \\
 0 & 0 & -1 & 0 & 2 & -1 & 0 \\
 0 & 0 & 0 & 0 & -1 & 2 & -1 \\
 0 & 0 & 0 & 0 & 0 & -1 & 1 \\
} & H7_{1} & \smat{
 2 & -1 & 0 & 0 & 0 & 0 & 0 \\
 -1 & 2 & -1 & 0 & 0 & 0 & 0 \\
 0 & -1 & 2 & -1 & -1 & 0 & 0 \\
 0 & 0 & -1 & 2 & 0 & 0 & 0 \\
 0 & 0 & -1 & 0 & 2 & -1 & 0 \\
 0 & 0 & 0 & 0 & -1 & 2 & -1 \\
 0 & 0 & 0 & 0 & 0 & -2 & 2 \\
} &
\end{matrix}
\]
\[\arraycolsep=1.5pt
\begin{matrix}
S8_{1} & \smat{
 2 & -1 & 0 & 0 & 0 & 0 & 0 & 0 \\
 -1 & 2 & -1 & 0 & 0 & 0 & 0 & 0 \\
 0 & -1 & 2 & -1 & -1 & 0 & 0 & 0 \\
 0 & 0 & -1 & 2 & 0 & 0 & 0 & 0 \\
 0 & 0 & -1 & 0 & 2 & -1 & 0 & 0 \\
 0 & 0 & 0 & 0 & -1 & 2 & -1 & 0 \\
 0 & 0 & 0 & 0 & 0 & -1 & 2 & -1 \\
 0 & 0 & 0 & 0 & 0 & 0 & -1 & 1 \\
} & H8_{2} & \smat{
 2 & -1 & 0 & 0 & 0 & 0 & 0 & 0 \\
 -1 & 2 & -1 & 0 & 0 & 0 & 0 & 0 \\
 0 & -1 & 2 & -1 & -1 & 0 & 0 & 0 \\
 0 & 0 & -1 & 2 & 0 & 0 & 0 & 0 \\
 0 & 0 & -1 & 0 & 2 & -1 & 0 & 0 \\
 0 & 0 & 0 & 0 & -1 & 2 & -1 & 0 \\
 0 & 0 & 0 & 0 & 0 & -1 & 2 & -1 \\
 0 & 0 & 0 & 0 & 0 & 0 & -2 & 2 \\
} &
\end{matrix}
\]
\[\arraycolsep=1.5pt
\begin{matrix}
S9_{1} & \smat{
 2 & -1 & 0 & 0 & 0 & 0 & 0 & 0 & 0 \\
 -1 & 2 & -1 & 0 & 0 & 0 & 0 & 0 & 0 \\
 0 & -1 & 2 & -1 & -1 & 0 & 0 & 0 & 0 \\
 0 & 0 & -1 & 2 & 0 & 0 & 0 & 0 & 0 \\
 0 & 0 & -1 & 0 & 2 & -1 & 0 & 0 & 0 \\
 0 & 0 & 0 & 0 & -1 & 2 & -1 & 0 & 0 \\
 0 & 0 & 0 & 0 & 0 & -1 & 2 & -1 & 0 \\
 0 & 0 & 0 & 0 & 0 & 0 & -1 & 2 & -1 \\
 0 & 0 & 0 & 0 & 0 & 0 & 0 & -1 & 1 \\
} & H9_{2} & \smat{
 2 & -1 & 0 & 0 & 0 & 0 & 0 & 0 & 0 \\
 -1 & 2 & -1 & 0 & 0 & 0 & 0 & 0 & 0 \\
 0 & -1 & 2 & -1 & -1 & 0 & 0 & 0 & 0 \\
 0 & 0 & -1 & 2 & 0 & 0 & 0 & 0 & 0 \\
 0 & 0 & -1 & 0 & 2 & -1 & 0 & 0 & 0 \\
 0 & 0 & 0 & 0 & -1 & 2 & -1 & 0 & 0 \\
 0 & 0 & 0 & 0 & 0 & -1 & 2 & -1 & 0 \\
 0 & 0 & 0 & 0 & 0 & 0 & -1 & 2 & -1 \\
 0 & 0 & 0 & 0 & 0 & 0 & 0 & -2 & 2 \\
} &
\end{matrix}
\]
\[\arraycolsep=1.5pt
\begin{matrix}
S10_{1} & \smat{
 2 & -1 & 0 & 0 & 0 & 0 & 0 & 0 & 0 & 0 \\
 -1 & 2 & -1 & 0 & 0 & 0 & 0 & 0 & 0 & 0 \\
 0 & -1 & 2 & -1 & -1 & 0 & 0 & 0 & 0 & 0 \\
 0 & 0 & -1 & 2 & 0 & 0 & 0 & 0 & 0 & 0 \\
 0 & 0 & -1 & 0 & 2 & -1 & 0 & 0 & 0 & 0 \\
 0 & 0 & 0 & 0 & -1 & 2 & -1 & 0 & 0 & 0 \\
 0 & 0 & 0 & 0 & 0 & -1 & 2 & -1 & 0 & 0 \\
 0 & 0 & 0 & 0 & 0 & 0 & -1 & 2 & -1 & 0 \\
 0 & 0 & 0 & 0 & 0 & 0 & 0 & -1 & 2 & -1 \\
 0 & 0 & 0 & 0 & 0 & 0 & 0 & 0 & -1 & 1 \\
} & H10_{2} & \smat{
 2 & -1 & 0 & 0 & 0 & 0 & 0 & 0 & 0 & 0 \\
 -1 & 2 & -1 & 0 & 0 & 0 & 0 & 0 & 0 & 0 \\
 0 & -1 & 2 & -1 & -1 & 0 & 0 & 0 & 0 & 0 \\
 0 & 0 & -1 & 2 & 0 & 0 & 0 & 0 & 0 & 0 \\
 0 & 0 & -1 & 0 & 2 & -1 & 0 & 0 & 0 & 0 \\
 0 & 0 & 0 & 0 & -1 & 2 & -1 & 0 & 0 & 0 \\
 0 & 0 & 0 & 0 & 0 & -1 & 2 & -1 & 0 & 0 \\
 0 & 0 & 0 & 0 & 0 & 0 & -1 & 2 & -1 & 0 \\
 0 & 0 & 0 & 0 & 0 & 0 & 0 & -1 & 2 & -1 \\
 0 & 0 & 0 & 0 & 0 & 0 & 0 & 0 & -2 & 2 \\
} &
\end{matrix}
\]

}


\section[The pairs ``almost affine Lie superalgebra longleftrightarrow hyperbolic Lie algebra'' for non-symmetrizable Cartan matrices]{The pairs ``almost affine Lie superalgebra $\boldsymbol{\longleftrightarrow}$ hyperbolic\\ Lie algebra'' for non-symmetrizable Cartan matrices}\label{section3}

This section is added for completeness; we do not know of any of its possible interpretations.

In this section, we desuperize non-symmetrizable Cartan matrices of almost affine Lie superalgebras. The matrices with multiple (in parentheses) ``superizations'':
\[
NH3_{25}(3), \ \ NH3_{29}(3), \ \ NH3_{85}(3).
\]

Total NS matrices: 36, total NH matrices admitting ``superization'': 30 of 96,{\tiny
\[\arraycolsep=1.5pt
\begin{matrix}
NS3_{1} & \smat{
 2 & -1 & -1 \\
 -1 & 1 & -1 \\
 -2 & -1 & 2 \\
} & NH3_{1} & \smat{
 2 & -1 & -1 \\
 -2 & 2 & -1 \\
 -2 & -2 & 2 \\
} & \{1,3,2\} & \qquad NS3_{2} & \smat{
 2 & -1 & -1 \\
 -1 & 1 & -1 \\
 -3 & -1 & 2 \\
} & NH3_{2} & \smat{
 2 & -1 & -1 \\
 -3 & 2 & -1 \\
 -2 & -2 & 2 \\
} & \{1,3,2\} \\ \\
NS3_{3} & \smat{
 2 & -1 & -1 \\
 -1 & 1 & -1 \\
 -4 & -1 & 2 \\
} & NH3_{3} & \smat{
 2 & -1 & -1 \\
 -4 & 2 & -1 \\
 -2 & -2 & 2 \\
} & \{1,3,2\} & \qquad NS3_{7} & \smat{
 2 & -1 & -1 \\
 -2 & 1 & -1 \\
 -3 & -1 & 2 \\
} & NH3_{15} & \smat{
 2 & -1 & -1 \\
 -3 & 2 & -1 \\
 -4 & -2 & 2 \\
} & \{1,3,2\} \\ \\
NS3_{8} & \smat{
 2 & -1 & -1 \\
 -2 & 1 & -1 \\
 -4 & -1 & 2 \\
} & NH3_{16} & \smat{
 2 & -1 & -1 \\
 -4 & 2 & -1 \\
 -4 & -2 & 2 \\
} & \{1,3,2\} & \qquad NS3_{9} & \smat{
 2 & -1 & -2 \\
 -2 & 1 & -1 \\
 -2 & -1 & 2 \\
} & NH3_{17} & \smat{
 2 & -2 & -1 \\
 -2 & 2 & -1 \\
 -4 & -2 & 2 \\
} & \{1,3,2\} \\ \\
NS3_{10} & \smat{
 2 & -1 & -1 \\
 -2 & 1 & -1 \\
 -1 & -1 & 2 \\
} & NH3_{18} & \smat{
 2 & -1 & -1 \\
 -1 & 2 & -1 \\
 -4 & -2 & 2 \\
} & \{1,3,2\} & \qquad NS3_{11} & \smat{
 2 & -1 & -2 \\
 -2 & 1 & -1 \\
 -1 & -1 & 2 \\
} & NH3_{19} & \smat{
 2 & -2 & -1 \\
 -1 & 2 & -1 \\
 -4 & -2 & 2 \\
} & \{1,3,2\} \\ \\
NS3_{12} & \smat{
 2 & -1 & -3 \\
 -2 & 1 & -1 \\
 -1 & -1 & 2 \\
} & NH3_{20} & \smat{
 2 & -3 & -1 \\
 -1 & 2 & -1 \\
 -4 & -2 & 2 \\
} & \{1,3,2\} & \qquad NS3_{13} & \smat{
 2 & -1 & -4 \\
 -2 & 1 & -1 \\
 -1 & -1 & 2 \\
} & NH3_{21} & \smat{
 2 & -4 & -1 \\
 -1 & 2 & -1 \\
 -4 & -2 & 2 \\
} & \{1,3,2\} \\ \\
NS3_{14} & \smat{
 2 & -2 & -1 \\
 -1 & 1 & -1 \\
 -2 & -1 & 2 \\
} & NH3_{22} & \smat{
 2 & -1 & -2 \\
 -2 & 2 & -1 \\
 -2 & -2 & 2 \\
} & \{1,3,2\} & \qquad NS3_{15} & \smat{
 2 & -2 & -1 \\
 -1 & 1 & -1 \\
 -3 & -1 & 2 \\
} & NH3_{23} & \smat{
 2 & -1 & -2 \\
 -3 & 2 & -1 \\
 -2 & -2 & 2 \\
} & \{1,3,2\} \\ \\
NS3_{16} & \smat{
 2 & -2 & -1 \\
 -1 & 1 & -1 \\
 -4 & -1 & 2 \\
} & NH3_{24} & \smat{
 2 & -1 & -2 \\
 -4 & 2 & -1 \\
 -2 & -2 & 2 \\
} & \{1,3,2\} & \qquad \fbox{!!!}~~NS3_{17} & \smat{
 2 & -2 & -2 \\
 -1 & 1 & -1 \\
 -2 & -1 & 2 \\
} & NH3_{25} & \smat{
 2 & -2 & -2 \\
 -2 & 2 & -1 \\
 -2 & -2 & 2 \\
} & \{1,3,2\} \\ \\
NS3_{18} & \smat{
 2 & -2 & -1 \\
 -1 & 1 & -1 \\
 -1 & -1 & 2 \\
} & NH3_{26} & \smat{
 2 & -1 & -2 \\
 -1 & 2 & -1 \\
 -2 & -2 & 2 \\
} & \{1,3,2\} & \qquad NS3_{20} & \smat{
 2 & -2 & -3 \\
 -1 & 1 & -1 \\
 -1 & -1 & 2 \\
} & NH3_{28} & \smat{
 2 & -3 & -2 \\
 -1 & 2 & -1 \\
 -2 & -2 & 2 \\
} & \{1,3,2\} \\ \\
\fbox{!!!}~~NS3_{21} & \smat{
 2 & -2 & -4 \\
 -1 & 1 & -1 \\
 -1 & -1 & 2 \\
} & NH3_{29} & \smat{
 2 & -4 & -2 \\
 -1 & 2 & -1 \\
 -2 & -2 & 2 \\
} & \{1,3,2\} & \qquad NS3_{22} & \smat{
 2 & -1 & -1 \\
 -2 & 1 & -2 \\
 -2 & -1 & 2 \\
} & NH3_{52} & \smat{
 2 & -4 & -4 \\
 -1 & 2 & -1 \\
 -1 & -2 & 2 \\
} & \{2,1,3\} \\ \\
NS3_{23} & \smat{
 2 & -1 & -1 \\
 -2 & 1 & -2 \\
 -3 & -1 & 2 \\
} & NH3_{80} & \smat{
 2 & -4 & -4 \\
 -1 & 2 & -1 \\
 -1 & -3 & 2 \\
} & \{2,1,3\} & \qquad NS3_{24} & \smat{
 2 & -1 & -1 \\
 -2 & 1 & -2 \\
 -4 & -1 & 2 \\
} & NH3_{81} & \smat{
 2 & -1 & -1 \\
 -4 & 2 & -1 \\
 -4 & -4 & 2 \\
} & \{1,3,2\} \\ \\
NS3_{27} & \smat{
 2 & -2 & -1 \\
 -1 & 1 & -2 \\
 -2 & -1 & 2 \\
} & NH3_{49} & \smat{
 2 & -2 & -4 \\
 -2 & 2 & -1 \\
 -1 & -2 & 2 \\
} & \{2,1,3\} & \qquad NS3_{28} & \smat{
 2 & -2 & -1 \\
 -1 & 1 & -2 \\
 -3 & -1 & 2 \\
} & NH3_{78} & \smat{
 2 & -2 & -4 \\
 -2 & 2 & -1 \\
 -1 & -3 & 2 \\
} & \{2,1,3\} \\ \\
NS3_{29} & \smat{
 2 & -2 & -1 \\
 -1 & 1 & -2 \\
 -4 & -1 & 2 \\
} & NH3_{84} & \smat{
 2 & -1 & -2 \\
 -4 & 2 & -1 \\
 -2 & -4 & 2 \\
} & \{1,3,2\} & \qquad \fbox{!!!}~~NS3_{30} & \smat{
 2 & -2 & -2 \\
 -1 & 1 & -2 \\
 -2 & -1 & 2 \\
} & NH3_{85} & \smat{
 2 & -2 & -2 \\
 -2 & 2 & -1 \\
 -2 & -4 & 2 \\
} & \{1,3,2\} \\ \\
NS3_{31} & \smat{
 2 & -2 & -1 \\
 -1 & 1 & -2 \\
 -1 & -1 & 2 \\
} & NH3_{86} & \smat{
 2 & -1 & -2 \\
 -1 & 2 & -1 \\
 -2 & -4 & 2 \\
} & \{1,3,2\} & \qquad \fbox{!!!}~~NS3_{32} & \smat{
 2 & -2 & -2 \\
 -1 & 1 & -2 \\
 -1 & -1 & 2 \\
} & NH3_{29} & \smat{
 2 & -4 & -2 \\
 -1 & 2 & -1 \\
 -2 & -2 & 2 \\
} & \{3,1,2\} \\ \\
NS3_{33} & \smat{
 2 & -2 & -3 \\
 -1 & 1 & -2 \\
 -1 & -1 & 2 \\
} & NH3_{68} & \smat{
 2 & -4 & -2 \\
 -1 & 2 & -1 \\
 -2 & -3 & 2 \\
} & \{3,1,2\} & \qquad \fbox{!!!}~~NS3_{35} & \smat{
 2 & -2 & -1 \\
 -1 & 1 & -1 \\
 -2 & -2 & 2 \\
} & NH3_{25} & \smat{
 2 & -2 & -2 \\
 -2 & 2 & -1 \\
 -2 & -2 & 2 \\
} & \{2,1,3\} \\ \\
NS3_{36} & \smat{
 2 & -2 & -1 \\\textsl{•}
 -1 & 1 & -1 \\
 -3 & -2 & 2 \\
} & NH3_{65} & \smat{
 2 & -2 & -2 \\
 -2 & 2 & -1 \\
 -2 & -3 & 2 \\
} & \{2,1,3\} & \qquad \fbox{!!!}~~NS3_{37} & \smat{
 2 & -2 & -1 \\
 -1 & 1 & -1 \\
 -4 & -2 & 2 \\
} & NH3_{85} & \smat{
 2 & -2 & -2 \\
 -2 & 2 & -1 \\
 -2 & -4 & 2 \\
} & \{2,1,3\} \\ \\
\fbox{!!!}~~NS3_{41} & \smat{
 2 & -1 & -1 \\
 -2 & 1 & -1 \\
 -1 & -1 & 1 \\
} & NH3_{29} & \smat{
 2 & -4 & -2 \\
 -1 & 2 & -1 \\
 -2 & -2 & 2 \\
} & \{2,1,3\} & \qquad \fbox{!!!}~~NS3_{42} & \smat{
 2 & -2 & -1 \\
 -1 & 1 & -1 \\
 -1 & -1 & 1 \\
} & NH3_{25} & \smat{
 2 & -2 & -2 \\
 -2 & 2 & -1 \\
 -2 & -2 & 2 \\
} & \{2,1,3\} \\ \\
\fbox{!!!}~~NS3_{44} & \smat{
 2 & -2 & -1 \\
 -1 & 1 & -1 \\
 -2 & -1 & 1 \\
} & NH3_{85} & \smat{
 2 & -2 & -2 \\
 -2 & 2 & -1 \\
 -2 & -4 & 2 \\
} & \{2,1,3\} & \qquad & & & &
\end{matrix}
\]
\[\arraycolsep=1.5pt
\begin{matrix}
NS4_{15} & \smat{
 2 & -1 & 0 & -2 \\
 -1 & 1 & -1 & 0 \\
 0 & -1 & 2 & -1 \\
 -1 & 0 & -1 & 2 \\
} & NH4_{20} & \smat{
 2 & -1 & 0 & -1 \\
 -2 & 2 & -1 & 0 \\
 0 & -2 & 2 & -2 \\
 -1 & 0 & -1 & 2 \\
} & \{2,3,4,1\} & \qquad NS4_{16} & \smat{
 2 & -1 & 0 & -1 \\
 -1 & 1 & -1 & 0 \\
 0 & -1 & 2 & -1 \\
 -2 & 0 & -1 & 2 \\
} & NH4_{21} & \smat{
 2 & -2 & 0 & -1 \\
 -1 & 2 & -1 & 0 \\
 0 & -2 & 2 & -2 \\
 -1 & 0 & -1 & 2 \\
} & \{2,3,4,1\} \\ \\
NS4_{18} & \smat{
 2 & -1 & 0 & -1 \\
 -1 & 1 & -1 & 0 \\
 0 & -1 & 2 & -2 \\
 -2 & 0 & -1 & 2 \\
} & NH4_{23} & \smat{
 2 & -2 & 0 & -1 \\
 -1 & 2 & -1 & 0 \\
 0 & -2 & 2 & -2 \\
 -2 & 0 & -1 & 2 \\
} & \{2,3,4,1\} & \qquad & & & &
\end{matrix}
\]
}

\subsection{Remark} For completeness, observe that among the simple finite-dimensional Lie superalgebras $\fg$ without Cartan matrix, there are 2 series such that $\fg_\ev$ is a~simple Lie algebra.
To describe such pairs $(\fg, \fg_\ev)$, recall (see, e.g., \cite{CCLL}) that $\fq(n)$, called the \textit{queer} Lie superalgebra, is another super analog of $\fgl(n)$, in addition to $\fgl(a|b)$. These super analogs, namely $\fgl(a|b)$ and $\fq(n)$, correspond, respectively, to the two cases of the Schur lemma: when the irreducible module has no invariant subsuperspaces and when it has no invariant subsuperspaces, but has an invariant subspace. In the latter case, $\fq(n)$ can be interpreted as preserving a~complex structure given by an \textit{odd} operator $J$ in an $(n|n)$-dimensional superspace~$V$. Having selected a basis of standard format in~$V$ so that the matrix of~$J$ is $\left(\begin{smallmatrix}0&1_n\\-1_n&0\end{smallmatrix}\right)$, we can realize the supermatrices of $\fq(n)$ in the form $(A,B):=\left(\begin{smallmatrix}A&B\\ B&A\end{smallmatrix}\right)$, where $A,B\in\fgl(n)$. The queer trace $\qtr\colon (A,B)\mapsto \tr B$ singles out a~queertraceless subalgebra $\fsq(n)$; set $\fpsq(n):=\fsq(n)/\Cee 1_{2n}$.

The Lie superalgebra $\fpe(n)$ preserving a~non-degenerate invariant odd symmetric bilinear form in an $(n|n)$-dimensional superspace is called \textit{periplectic}, as A.~Weil suggested. Let $\str$ denote the \textit{supertrace} --- a~functional on the Lie superalgebra $\fg$ that vanishes on $\fg'$. Set
\[
\fspe(n):=\{X\in \fpe(n)\mid \str X=0\}.
\]

The two series described above are, for $n\geq 3$ (here $\Pi$ is the inversion of parity functor):
\[
\begin{array}{l}
\fpsq(n)\text{~~with $\fpsq(n)_\ev\simeq\fsl(n)$ and $\fpsq(n)_\od\simeq\Pi(\fsl(n))$};\\
\fspe(n)\text{~~with $\fspe(n)_\ev\simeq\fsl(n)=\fsl(V)$ and $\fspe(n)_\od\simeq\begin{cases}\Pi(\Lambda^2(V)\oplus S^2(V^*))&\text{or}\\
\Pi(\Lambda^2(V^*)\oplus S^2(V)).\\
\end{cases}$}
\end{array}
\]

\section{Open question}\label{opq} We observed that the Weyl chamber of a given hyperbolic Lie algebra $\fg$ and the Weyl chamber of any of the ``superizations'' of $\fg$ are one and the same cone.

Indeed, the ``superization'' \eqref{step1} multiplies one of the simple roots by~2 and leaves the Cartan subalgebra $\fh$ and the other simple roots as they were. The Weyl chamber
\[
C=\{h\in\fh_{\Ree}\mid \alpha(h)>0 \text{ for every simple root } \alpha\}
\]
depends on the simple roots only up to a~positive factor each, since $\alpha(h)>0\Longleftrightarrow 2\alpha(h)>0$; hence the two chambers coincide. For the same reason, $\alpha$ and $2\alpha$ vanish on one and the same hyperplane and the reflections in them are equal, so the Lie algebra and its ``superization'' have not only the same chamber but the same system of walls and the same Weyl group; the walls just carry roots of both parities now. What does differ is the root system itself, which for the ``superization'' is non-reduced: together with some roots $\alpha$ of $\fg$, the Lie superalgebra has the roots $2\alpha$, as $\fosp(1|2n)$ has both $\pm\varepsilon_i$ and $\pm2\varepsilon_i$. The billiard table is therefore literally the same table, and this is why the tracks coincide.

How does this ``superization'' of the conventional billiard affect the model describing the cosmological billiard, cf.\ \cite{BH, KhK1, BKhL} and \cite{DSp, BS, KKN}?

\subsection*{Acknowledgements}
D.L.\ was supported by the grant AD 065 NYUAD in 2021.
We heartily thank the referees for the careful refereeing, criticism, and several
very helpful constructive remarks.
We acknowledge the letter (2024-09-23) from Lisa Carbone to D.L.: ``I have seen your recent paper `On a~hidden supersymmetry of cosmological billiards.' Some of the results were previously worked out in my paper \cite{Cetal}.
I think it would be appropriate for you to add a citation.'' So we added it for the reader (and all the co-authors of \cite{Cetal}) to enjoy.


\pdfbookmark[1]{References}{ref}
\LastPageEnding

\end{document}